\documentclass[12pt]{iopart}
\usepackage{iopams}
\usepackage{graphicx}
\begin{document}
%
%

\title{A spatial correlation analysis for a toroidal universe}

\author{Ralf Aurich$^1$}

\address{$^1$Institut f\"ur Theoretische Physik, Universit\"at Ulm,\\
Albert-Einstein-Allee 11, D-89069 Ulm, Germany}

\begin{abstract}
The spatial cross-correlation function $\xi_{\hbox{\scriptsize C}}$ recently
introduced by Roukema et al.\ \cite{Roukema_et_al_2008a}
is applied to the equilateral toroidal topology of the universe.
Several CMB maps based on the WMAP 5yr data are analysed
and a small likelihood in favour of a torus cell is revealed.
The results are compared to ten $\Lambda$CDM simulations
which point to a high false positive rate of the spatial-correlation-function method
such that a firm conclusion cannot be drawn.
\end{abstract}



\section{Introduction}

The cosmic microwave background radiation (CMB) provides not only
the information to constrain the cosmological parameters,
but also has the potential to reveal the global structure of the Universe,
i.\,e.\ its topology \cite{Lachieze-Rey_Luminet_1995,Starkman_1998,%
Luminet_Roukema_1999,Levin_2002,Reboucas_Gomero_2004,Luminet_2008}.
One method to reveal the topology is the search for the so-called
circles-in-the-sky signature \cite{Cornish_Spergel_Starkman_1998b}.
The CMB descends from the surface of last scattering (SLS) which is the
spherical surface around an observer at which matter is seen to recombine.
In the case of a multiply connected universe this sphere intersects with its
``copies'' due to the group of deck transformations defining the topology.
Since the intersection of two spheres is a circle, one obtains pairs of circles
along which the temperature fluctuations are identical, or at least,
if we could receive the CMB as a pure signal proportional to the
gravitational field on the SLS.
Modifications due to the Doppler and the integrated Sachs-Wolfe effect
are different on two paired circles thus preventing identical
temperature fluctuations along paired circles.
A further complication arises from the uncertainties of foreground emissions
which additionally influence the observed CMB.
Applications of the circles-in-the-sky signature to the WMAP data
\cite{Hinshaw_et_al_2008} can be found in
\cite{Cornish_Spergel_Starkman_Komatsu_2003,%
Roukema_et_al_2004,Aurich_Lustig_Steiner_2005b,%
Key_Cornish_Spergel_Starkman_2007,Lew_Roukema_2008}.

The circles-in-the-sky method utilizes only the one-dimensional
pixel information along paired circles.
To improve the topological signal it is proposed in \cite{Roukema_et_al_2008a}
that instead of analyzing correlations along the circles,
one should define two {\it spatial} correlation functions
$\xi_{\hbox{\scriptsize A}}$ and $\xi_{\hbox{\scriptsize C}}$
which also depend on pixels not lying on paired circles.
Then the comparison of $\xi_{\hbox{\scriptsize A}}$ with
$\xi_{\hbox{\scriptsize C}}$
provides the topological signature.
The idea is the following:
The topology is defined by the group of deck transformations $\Gamma$,
i.\,e.\ a discrete subgroup of isometries without fixed points. 
All spatial points $q_1, q_2 \in{\cal M}^{(3)}$ obeying
$q_1 = \gamma(q_2)$ for $\gamma \in \Gamma$ are identified,
i.\,e.\ the spatial comoving space section ${\cal M}^{(3)}$ is tesselated by
considering the quotient ${\cal M}^{(3)}/\Gamma$.
Define now the set $\tilde\Gamma$ which contains the elements
of $\Gamma$ but the identity removed.
With respect to this $\tilde\Gamma$ the distance $d_{\hbox{\scriptsize topo}}$
between two points $q_1, q_2 \in{\cal M}^{(3)}$ is defined as
\begin{equation}
\label{Eq:Distance_topo}
d_{\hbox{\scriptsize topo}}(q_1,q_2) \; := \;
\min_{\gamma\in\tilde\Gamma} d(q_1,\gamma(q_2))
\hspace{10pt} ,
\end{equation}
where $d(q,q')$ is the usual spatial comoving distance
in the universal covering space ${\cal M}^{(3)}$.
Note, that for sufficiently nearby points $q_1$ and $q_2$
one has $d_{\hbox{\scriptsize topo}}(q_1,q_2)>d(q_1,q_2)$
since the identity is not contained in $\tilde\Gamma$.
The reverse applies for sufficiently far separated points $q_1$ and $q_2$.
Now the two spatial correlation functions
$\xi_{\hbox{\scriptsize A}}$ and $\xi_{\hbox{\scriptsize C}}$
can be defined as
\begin{equation}
\xi_{\hbox{\scriptsize A}}(r) \; := \;
\left<\, \delta T(q_1)  \, \delta T(q_2) \,\right>
\hspace{10pt} \hbox{with}  \hspace{10pt}
r = d(q_1,q_2)
\end{equation}
for the auto-correlation function and
\begin{equation}
\xi_{\hbox{\scriptsize C}}(r) \; := \;
\left<\, \delta T(q_1)  \, \delta T(q_2) \,\right>
\hspace{10pt} \hbox{with}  \hspace{10pt}
r = d_{\hbox{\scriptsize topo}}(q_1,q_2)
\end{equation}
for the cross-correlation function.
Here, $\delta T(q)$ denotes the temperature fluctuation of the CMB
in the direction $\hat n$ corresponding to $q$
and $\left<\, \dots \,\right>$ the averaging over the pixels
with $r = d(q_1,q_2)$ or $r = d_{\hbox{\scriptsize topo}}(q_1,q_2)$.
In the practical analysis a binning with respect to $r$ is necessary
due to the pixelized sky maps.

The auto-correlation function $\xi_{\hbox{\scriptsize A}}(r)$ is the usual
spatial correlation function which has a pronounced peak at $r=0$.
In the cross-correlation function $\xi_{\hbox{\scriptsize C}}(r)$ this peak has
to be produced by points far separated on the sky
which are topologically adjacent due to $d_{\hbox{\scriptsize topo}}$.
One has detected the correct group $\Gamma$ when both correlation functions
$\xi_{\hbox{\scriptsize A}}(r)$ and $\xi_{\hbox{\scriptsize C}}(r)$
are identical,
if it were not for the same modifications of the CMB
which also affect the circles-in-the-sky method.
The main advantage over the circles-in-the-sky method is
that the correlations $\xi_{\hbox{\scriptsize A}}$ and
$\xi_{\hbox{\scriptsize C}}$ are computed from a much larger
set of pixels which levels out the modifications to some degree.
On the other hand, since the spatial quantities $\xi_{\hbox{\scriptsize A}}$
and $\xi_{\hbox{\scriptsize C}}$ are three-dimensional measures
which are only sampled from the two-dimensional SLS,
one cannot expect $\xi_{\hbox{\scriptsize A}}$ and
$\xi_{\hbox{\scriptsize C}}$ to be identical for statistical reasons.

\begin{figure}[ttt]
\begin{center}
\hspace*{0pt}\begin{minipage}{12cm}
\vspace*{-5pt}\includegraphics[width=12.0cm]{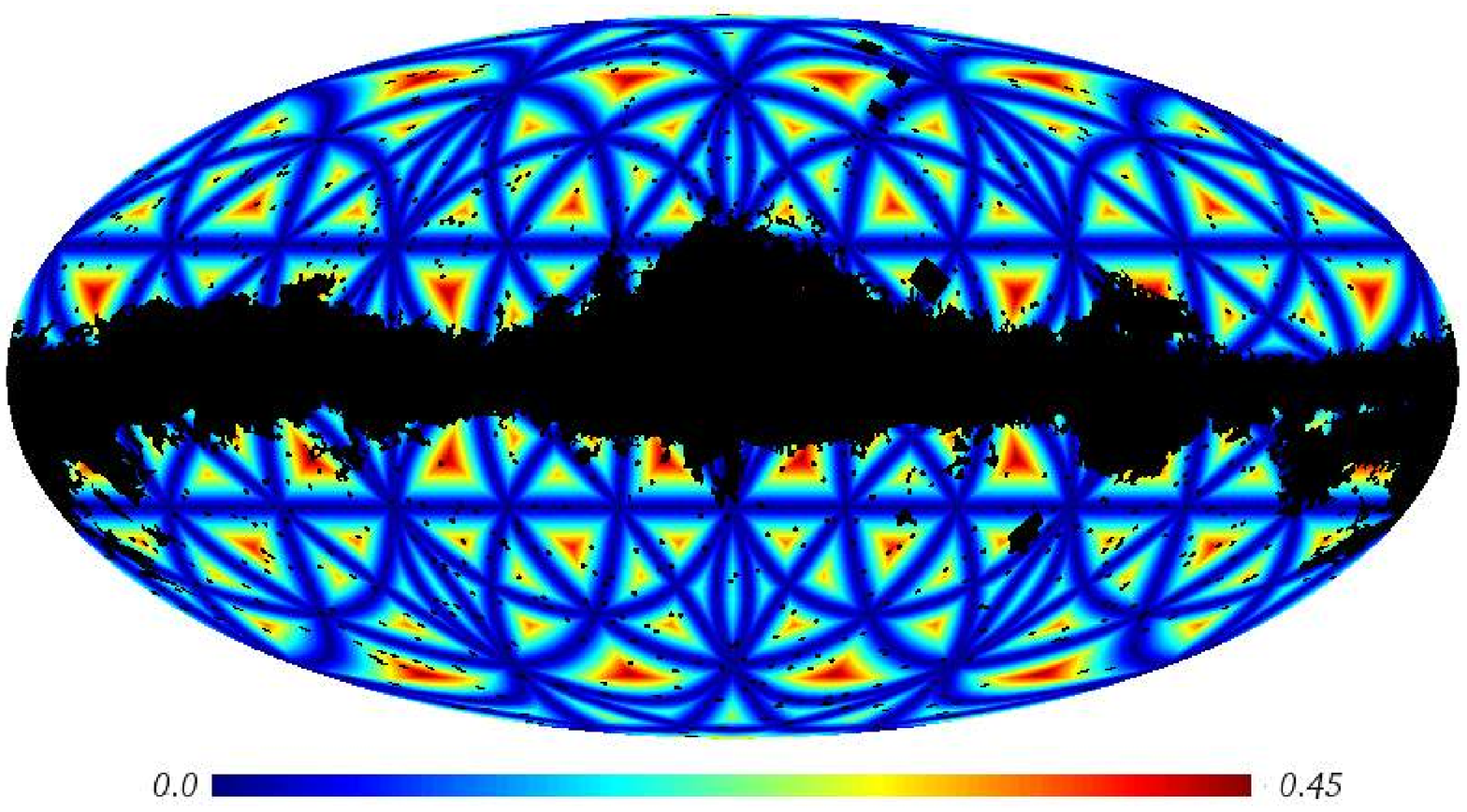}
\vspace*{-5pt}\includegraphics[width=12.0cm]{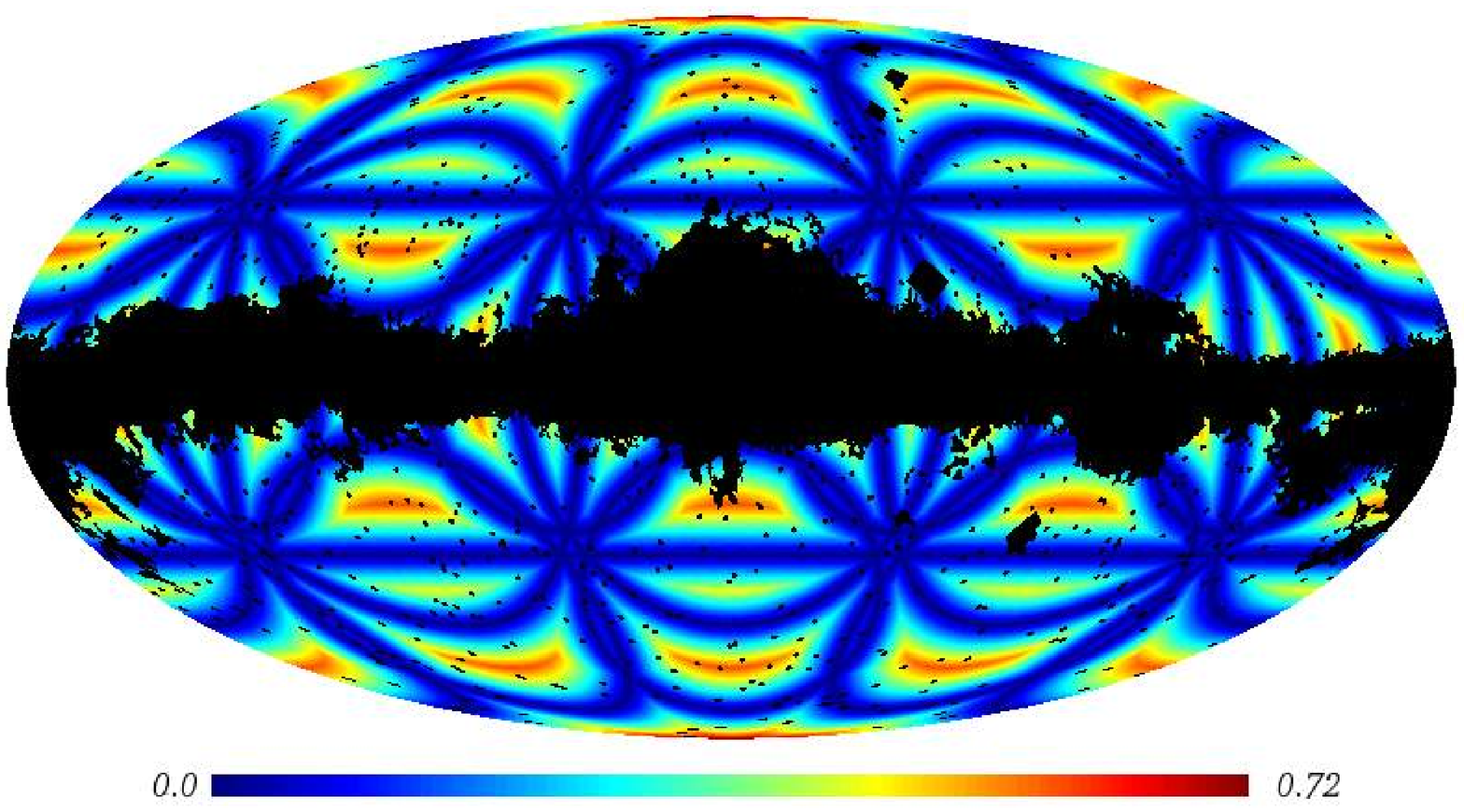}
\end{minipage}
\put(-360,170){a) \; \; $L=3$}
\put(-360,-20){b) \; \; $L=4$}
\vspace*{-10pt}
\end{center}
\caption{\label{Fig:MinDist_Torus}
The distance $d_{\hbox{\scriptsize min}}(q)$,
eq.\,(\ref{Eq:Distance_min}), is shown for the equilateral toroidal universe
with side lengths $L=3$ and $L=4$ using the Mollweide projection.
The paired circles correspond to $d_{\hbox{\scriptsize min}}=0$,
i.\,e.\ the dark blue regions.
The black regions show the pixels excluded by the KQ75 mask
used throughout this paper.
}
\end{figure}

To illustrate the difference between the two methods
Fig.\,\ref{Fig:MinDist_Torus} displays the distance
\begin{equation}
\label{Eq:Distance_min}
d_{\hbox{\scriptsize min}}(q) \; := \;
\min_{q'} \; d_{\hbox{\scriptsize topo}}(q,q')
\end{equation}
for the toroidal topology
where all three side lengths are either equal to $L=3$ or $L=4$
(in units of the Hubble length $L_{\hbox{\scriptsize H}} = c/H_0$,
see section \ref{Sec:MCMC})
and the symmetry axis pointing to the north pole.
Note that the scale in the two panels of Fig.\,\ref{Fig:MinDist_Torus} is different.
In the case of $L=3$ the maximal value of $d_{\hbox{\scriptsize min}}(q)$
is only 0.45, whereas the case $L=4$ possesses the larger bound 0.72.
This demonstrates that the distribution of $d_{\hbox{\scriptsize min}}(q)$
extends over an increasingly larger range with increasing values of $L$
since the elements $\gamma\in\tilde\Gamma$ shift a given point over
a correspondingly larger distance.
The number of pixels used to discriminate the topology for a given orientation
is much larger for the spatial-correlation-function method than
for the circles-in-the-sky method which utilizes only the pixels
along the circles with $d_{\hbox{\scriptsize min}}=0$.
In this work all pixels inside the KQ75 mask
\cite{Hinshaw_et_al_2008,Gold_et_al_2008} are excluded
which are contaminated by the milky way and other foreground sources.
These excluded domains are shown in black in Fig.\,\ref{Fig:MinDist_Torus}.

\begin{figure}[htb]
\begin{center}
\begin{minipage}{10cm}
\vspace*{-25pt}\includegraphics[width=10.0cm]{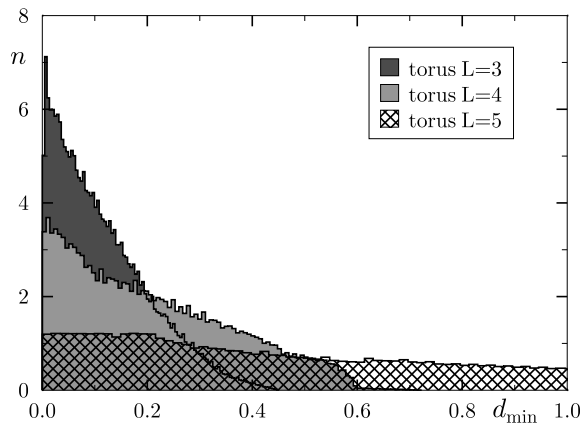}
\end{minipage}
\vspace*{-25pt}
\end{center}
\caption{\label{Fig:Hist_d_topo}
The distribution of the number of pixel pairs in dependence on
$d_{\hbox{\scriptsize min}}$ is shown for
three different sizes of the toroidal cell.
For the model with side length $L=3$, 4 and 5 all points on the SLS have a
largest distance  $d_{\hbox{\scriptsize min}}$ of 0.45, 0.72 and 1.66, respectively.
With decreasing torus size the number of pixel pairs with
$d_{\hbox{\scriptsize min}}<0.2$ increases significantly.
}
\end{figure}

The spatial-correlation-function method works best
if there is a large number of pixel pairs with small topological distances
$d_{\hbox{\scriptsize topo}}$,
since the peak of the correlation function at $r=0$ can then be tested
with a good statistical significance.
In units of the Hubble length one needs the correlation functions for $r<1$.
For this reason Fig.\,\ref{Fig:Hist_d_topo} presents the distribution
of the number of pixel pairs with respect to $d_{\hbox{\scriptsize min}}$
for three different sizes of the toroidal fundamental cell.
As noted above, with an increasing topological scale $L$,
the maximal value $d_{\hbox{\scriptsize min}}$ increases
since the points are topologically connected over ever larger distances.
A large number of pixel pairs close to $d{\hbox{\scriptsize min}}=0$ is observed
only for sufficiently small tori.
Since $d_{\hbox{\scriptsize min}}$ is the lower bound of the distance
for which the data are sampled for the estimation of the correlation functions,
this implies that the statistical significance for the estimates of
$\xi_{\hbox{\scriptsize C}}(r)$ for small values of $r$ diminishes with increasing $L$.
Thus for too large fundamental cells, the spatial-correlation-function method has the
same problems as the circles-in-the-sky method.
In the latter case the number and the sizes of paired circles declines with increasing
side length $L$ until there are none at all.
For $L=3$, 4, and 5 there are 16, 9, and 3
pairs of matched circles.
In the following the analysis is restricted to $L<5$.

In Roukema et al.\ \cite{Roukema_et_al_2008a} the 
spatial-correlation-function method is applied to the
Poincar\'e dodecahedron and a special orientation is found
for which an enhanced likelihood occurs.
In this paper, the flat toroidal topology with the additional restriction
that all side lengths are equal, i.\,e.\ the cubic topology, is in the focus
in order to see how this topology behaves using the
method of \cite{Roukema_et_al_2008a}.

\section{The Markov Chain Monte Carlo method for the toroidal universe}
\label{Sec:MCMC}

The toroidal topology is defined on the universal covering space
${\cal M}^{(3)}= \mathbb{R}^3$ on which the
points $q=(x,y,z)\in \mathbb{R}^3$ and $q'=(x',y',z')\in \mathbb{R}^3$ are identified by
\begin{equation}
q' \; = \; (x+n_x L_x, \; y+n_y L_y , \; z+n_z L_z)
\hspace{5pt} \hbox{ with } \hspace{5pt}
(n_x,n_y,n_z) \in \mathbb{Z}^3
\hspace{10pt} .
\end{equation}
Since the group $\Gamma$ is infinite,
a finite subset of $\tilde \Gamma$ has to be chosen for
the computation of the distance $d_{\hbox{\scriptsize topo}}(q_1,q_2)$
in eq.\,(\ref{Eq:Distance_topo}).
Here, the subset is restricted to the group elements
which lead to copies of the fundamental cell directly adjacent
over a face, an edge or a corner,
i.\,e.\ to the elements with $|n_x|\leq 1$, $|n_y|\leq 1$, $|n_z|\leq 1$
and excluding $(n_x,n_y,n_z)= (0,0,0)$, of course.
The minimum in eq.\,(\ref{Eq:Distance_topo}) is thus determined
over 26 group elements.
As already stated, only an equilateral toroidal cell is considered in the following,
i.\,e.\ $L=L_x=L_y=L_z$.
For a general toroidal topology, these considerations have to be modified
in an obvious way.
The lengths $L$ are given in units of the Hubble length
$L_{\hbox{\scriptsize H}} = c/H_0$.
For the conversion of the CMB correlation to the spatial correlation,
the distance to the SLS is required, which in turn is determined by the
chosen cosmological parameters.
For the $\Lambda$CDM model one gets from Table 2 in \cite{Spergel_et_al_2007}
the cosmological parameters based on all astronomical
observations, see their column ``3 Year + ALL Mean'', i.\,e.\
$\Omega_{\hbox{\scriptsize b}} = 0.044$,
$\Omega_{\hbox{\scriptsize cdm}} = 0.223$,
$\Omega_\Lambda = 0.733$, $h=0.704$, $n_s = 0.947$, $\tau=0.073$.
For these values the distance to the SLS is
$L_{\hbox{\scriptsize SLS}} = \Delta \eta L_{\hbox{\scriptsize H}} \simeq 14.2 \hbox{Gpc}$
where $\Delta \eta =\eta_0-\eta_{\hbox{\scriptsize SLS}} = 3.329$
($\eta$ is the conformal time).
For other sets of cosmological parameters,
one has to rescale the topological length $L$ with respect to the new value
of $\Delta \eta$.

As in \cite{Roukema_et_al_2008a} the Markov Chain Monte Carlo (MCMC) method
with the Metropolis algorithm \cite{Metropolis_Rosenbluth_Rosenbluth_Teller_Teller_1953}
is used in order to determine the most likely side length $L$ together with the
orientation of the fundamental cell defined by the three Euler angles $(\alpha, \beta, \gamma)$.
This leads to a four parameter $(L, \alpha, \beta, \gamma)$ search.
The Metropolis algorithm requires the specification of a probability estimator $P$
which determines the probability
$\min(1,P_{\hbox{\scriptsize new}}/P_{\hbox{\scriptsize old}})$
with which a new state is accepted in the Markov chain.
In \cite{Roukema_et_al_2008a} this probability is chosen as
\begin{equation}
\label{Eq:P_Roukema}
P \; := \; \prod_{i=1}^n \, \cases{
e^{-\frac{|\xi_{\hbox{\scriptsize C}}(i)-\xi_{\hbox{\scriptsize A}}(i)|^2}
{2\sigma_i^2}}
& for  $\xi_{\hbox{\scriptsize C}}(i) \leq \xi_{\hbox{\scriptsize A}}(i)$ \\
1 + 0.01 \frac{\xi_{\hbox{\scriptsize C}}(i)-\xi_{\hbox{\scriptsize A}}(i)}
{\xi_{\hbox{\scriptsize A}}(i)}
& for $\xi_{\hbox{\scriptsize C}}(i) \geq \xi_{\hbox{\scriptsize A}}(i)$
}
\hspace{10pt} ,
\end{equation}
with
\begin{equation}
\label{Eq:sigma_Roukema}
\sigma_i := \frac 12 \xi_{\hbox{\scriptsize A}}(i) \sqrt{\frac{N_n}{N_i}}
\hspace{10pt} .
\end{equation}
Here, the index $i$ runs over the bins for which the values of
$\xi_{\hbox{\scriptsize C}}$ and $\xi_{\hbox{\scriptsize A}}$
are sampled and $N_i$ denotes the number of pixel pairs contributing to bin $i$.
See ref.\ \cite{Roukema_et_al_2008a} for details and a discussion motivating this choice.
Since $\sigma_i$ is larger for bins having fewer entries $N_i$,
the probability $P$ weights the bins according to their statistical significance.
The smaller the distance $d_{\hbox{\scriptsize topo}}$ is,
the smaller is also the number of pixel pairs contributing to the corresponding bin.
However, it is just the behaviour at small distances,
i.\,e.\ the peak at $r=0$ in $\xi_{\hbox{\scriptsize A}}(r)$,
which has to be reproduced by $\xi_{\hbox{\scriptsize C}}(r)$
if the supposed group $\Gamma$ matches the true one.
For this reason a second probability $\hat P$ is used in this paper,
which emphasizes the peak at $r=0$ by using instead of the above $\sigma_i$,
eq.\,(\ref{Eq:sigma_Roukema}),
the following alternative
\begin{equation}
\label{Eq:sigma_peak}
\hat \sigma_i \; := \; \bar\xi_{\hbox{\scriptsize A}} \,
\sqrt{\frac{\bar\xi_{\hbox{\scriptsize A}}}{\xi_{\hbox{\scriptsize A}}(i)}}
\hspace{10pt} \hbox{ with } \hspace{10pt}
\bar\xi_{\hbox{\scriptsize A}} \; := \; \frac 1n \, \sum_{i=1}^n \xi_{\hbox{\scriptsize A}}(i)
\hspace{10pt} .
\end{equation}
For large values of $\xi_{\hbox{\scriptsize A}}(i)$ a small $\hat \sigma_i$
is obtained thus selecting models producing strong correlations at $r=0$.
This probability is complementary  to the former one.
It, however, ignores the number of entries on which the value of the
bin $i$ is based.

\begin{figure}[ttt]
\begin{center}
\begin{minipage}{10cm}
\vspace*{-25pt}\includegraphics[width=10.0cm]{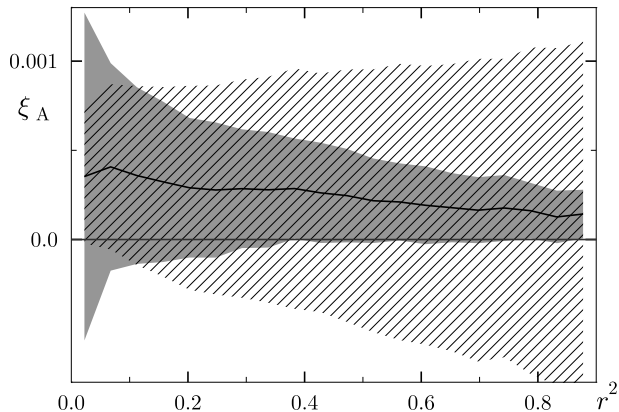}
\end{minipage}
\vspace*{-25pt}
\end{center}
\caption{\label{Fig:xi_error_bar}
The different behaviour of $\sigma$, eq.\,(\ref{Eq:sigma_Roukema}),
and $\hat \sigma$, eq.\,(\ref{Eq:sigma_peak}), is illustrated.
The grey region shows $\xi_{\hbox{\scriptsize A}}\pm\sigma$
which increases with with decreasing distances $r$.
The reverse behaviour is displayed by $\xi_{\hbox{\scriptsize A}}\pm\hat\sigma$
shown as the hatched area.
}
\end{figure}

Since the number of entries into a bin increases linearly for an equidistant
binning in $r$ for $\xi_{\hbox{\scriptsize A}}(i)$,
an equidistant binning with respect to $r^2$ is used in this paper.
This leads to $N_i$ of the same order.
For such a binning the complementary behaviour of $\sigma$ and $\hat \sigma$
is shown in Fig.\,\ref{Fig:xi_error_bar}.
The sky maps are in the HEALPix format
\cite{Gorski_Hivon_Banday_Wandelt_Hansen_Reinecke_Bartelmann_2005}
and are analysed for $n_{\hbox{\scriptsize side}} = 128$.
Since these maps have $12 n_{\hbox{\scriptsize side}}^2$ pixels,
only every 10th pixel is actually used in order to keep the
numerical effort manageable.
Without using a mask this leads to $N_{\hbox{\scriptsize pix}} = 19661$ pixels,
and using the KQ75 mask which cuts out 28\% of the area,
$N_{\hbox{\scriptsize pix}} = 13534$ pixels contribute
to the computation of $\xi_{\hbox{\scriptsize A}}$ and $\xi_{\hbox{\scriptsize C}}$.
The computation of the spatial correlation function is based on
$\frac 12 N_{\hbox{\scriptsize pix}} (N_{\hbox{\scriptsize pix}}-1)$
pixel pairs, which determine the increase of the numerical effort
with respect to the number of pixels $N_{\hbox{\scriptsize pix}}$.
It takes 1-2 weeks of cpu time on state of the art cpus to generate a
Markov chain with 20\,000 states for the above values.
In the likelihood (\ref{Eq:P_Roukema}) $n=20$ equidistant bins are used from
$r^2=0$ to $r^2=0.9$.
This binning is a compromise in order to resolve in a sufficient way the
structure of the spatial correlation functions and
to get enough entries into the bins for a statistical significant estimate.
Furthermore, the binning should not be finer than the corresponding resolution
of the HEALPix map,
since distances which are not resolved by the sky map cannot be
differentiated between adjacent bins. 
For $n_{\hbox{\scriptsize side}} = 128$  the pixel size is
$\theta_{\hbox{\scriptsize pix}} = 27.5'$
\cite{Gorski_Hivon_Banday_Wandelt_Hansen_Reinecke_Bartelmann_2005}
which translates into a spatial pixel size of $\Delta r \simeq 0.0266$.
Two points within two adjacent pixels can thus have a distance between 0
and $2\Delta r \simeq 0.053$;
thus $n \simeq 20$ represents the maximal useful binning.

A MCMC run is started by randomly generating an initial point
$(L_i, \alpha_i, \beta_i, \gamma_i)$ in the parameter space.
The initial values of $L_i$ are uniformly chosen from the interval $L_i\in [3.15, 3.45]$. 
By using the Metropolis algorithm,
the Markov chain is not restricted to this interval, however.
The initial Euler angles are uniformly chosen from the interval
$[0, \frac \pi 2]$.
Because of the symmetry of the toroidal fundamental cell,
there is some redundancy
in the sense that the set of Euler angles does not uniquely
define the orientation,
i.\,e.\ several sets of Euler angles can give the same values for the
Galactic coordinates of the symmetry axes of the fundamental cell.
In the following the symmetry axes denote the directions
in which one arrives at the centre of the nearest copies of the torus cell
by a shift of length $L$, i.\,e.\ the face-to-face neighbours.

\begin{figure}[ttt]
\begin{center}
\begin{minipage}{10cm}
\includegraphics[width=12.8cm]{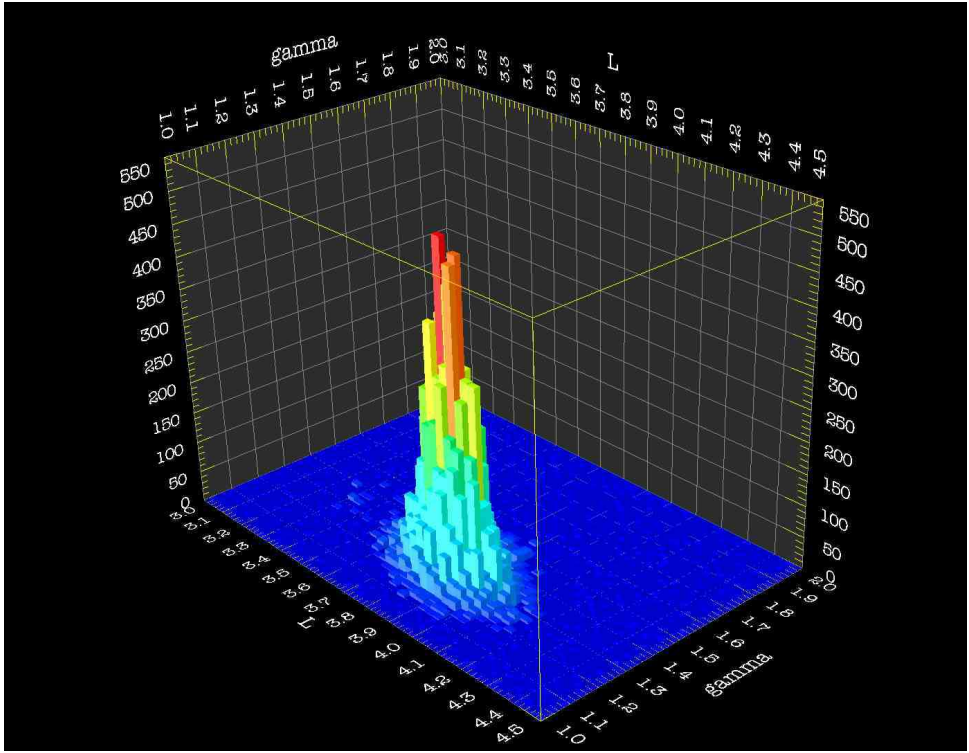}
\end{minipage}
\end{center}
\caption{\label{Fig:mcmc_ilc_distribution_L_gamma}
The distribution of the states of a Markov chain obtained from
the ILC 5yr map with the KQ75 mask is shown
in dependence on the length $L$ and the Euler angle $\gamma$.
The likelihood function with $\sigma$, eq.\,(\ref{Eq:sigma_Roukema}),
has been used.
}
\end{figure}

\section{The spatial-correlation-function signature for the toroidal universe}

Applying the MCMC method described in Section \ref{Sec:MCMC} to the cubic topology
one observes that, after the initial ``burn in'', most Markov chains occupy
a region in the parameter space corresponding to a side length $L\simeq 3.84$
using the WMAP-ILC map based on the 5yr data \cite{Hinshaw_et_al_2008}
outside the KQ75 mask.
Fig.\,\ref{Fig:mcmc_ilc_distribution_L_gamma} shows the distribution of a Markov chain
with respect to the two parameters $L$ and $\gamma$
ignoring $\alpha$ and $\beta$.
The distribution is localized around a small region in the parameter space
with Euler angles corresponding to the Galactical coordinates of
the toroidal symmetry axes given in Table \ref{Tab:Galactic_coordinates}.
However, this nice behaviour is not observed for all Markov chains.
If the initial length $L_i$ is too large, i.\,e.\ above 4.75, the Markov chains drift towards
ever larger values of $L$.
This is due to the decreasing statistical significance with increasing size of the
fundamental cell.
As shown in Fig.\,\ref{Fig:Hist_d_topo}
the number of pixels $N_r := \# \{ q | d_{\hbox{\scriptsize min}}(q)<r\}$,
which give information about the small distance behaviour
of the cross-correlation function $\xi_{\hbox{\scriptsize C}}(r)$,
decreases with increasing values of $L$.
Thus, the probability that the few remaining pixels generate by chance a peak
at $r=0$ in $\xi_{\hbox{\scriptsize C}}(r)$ rises with increasing values of $L$.
This unsatisfactory behaviour might be remedied by choosing another
probability function as (\ref{Eq:P_Roukema})
which has then to depend on $L$.
This topic will be addressed in a future study.

\begin{figure}[ttt]
\begin{center}
\begin{minipage}{10cm}
\vspace*{-25pt}\includegraphics[width=10.0cm]{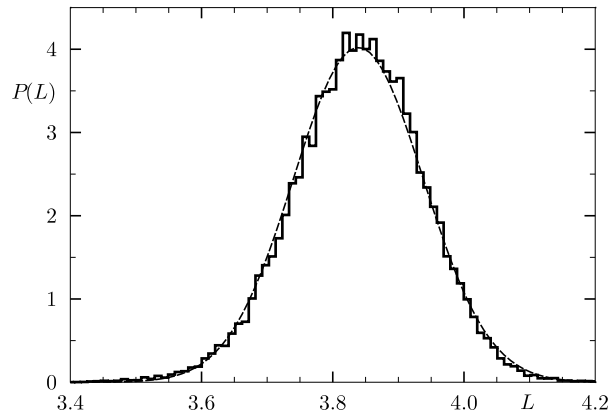}
\end{minipage}
\vspace*{-25pt}
\end{center}
\caption{\label{Fig:mcmc_distribution_L}
The normalized distribution of the states of ten Markov chains obtained from
the ILC 5yr map with the KQ75 mask is shown
in dependence on the length $L$.
The likelihood function with $\sigma$, eq.\,(\ref{Eq:sigma_Roukema}),
has been used.
The states of ten chains are merged so that the shown histogram
is based on 200\,000 states.
These chains lead to $L = 3.84 \pm 0.10$.
The dashed curve displays the Gaussian parameterized by these values.
}
\end{figure}

\begin{figure}[ttt]
\begin{center}
\begin{minipage}{10cm}
\vspace*{-5pt}\includegraphics[width=10.0cm]{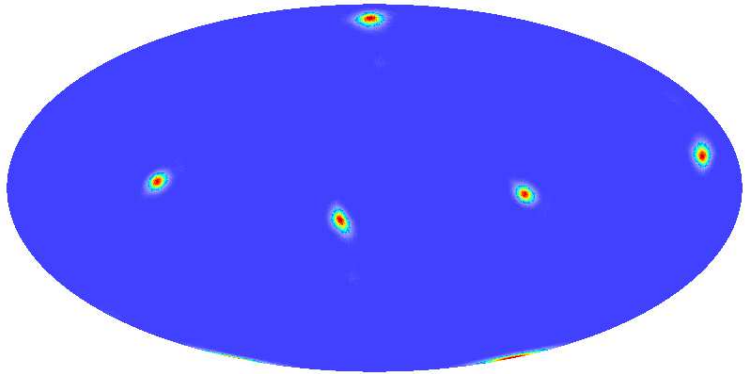}
\end{minipage}
\vspace*{-5pt}
\end{center}
\caption{\label{Fig:mcmc_distribution_angular}
The orientations of the symmetry axes of the Markov chains used in
Fig.\,\ref{Fig:mcmc_distribution_L} are shown on the sky in the Mollweide projection.
}
\end{figure}

\begin{figure}[ttt]
\begin{center}
\begin{minipage}{10cm}
\hspace*{-70pt}\includegraphics[width=16.0cm]{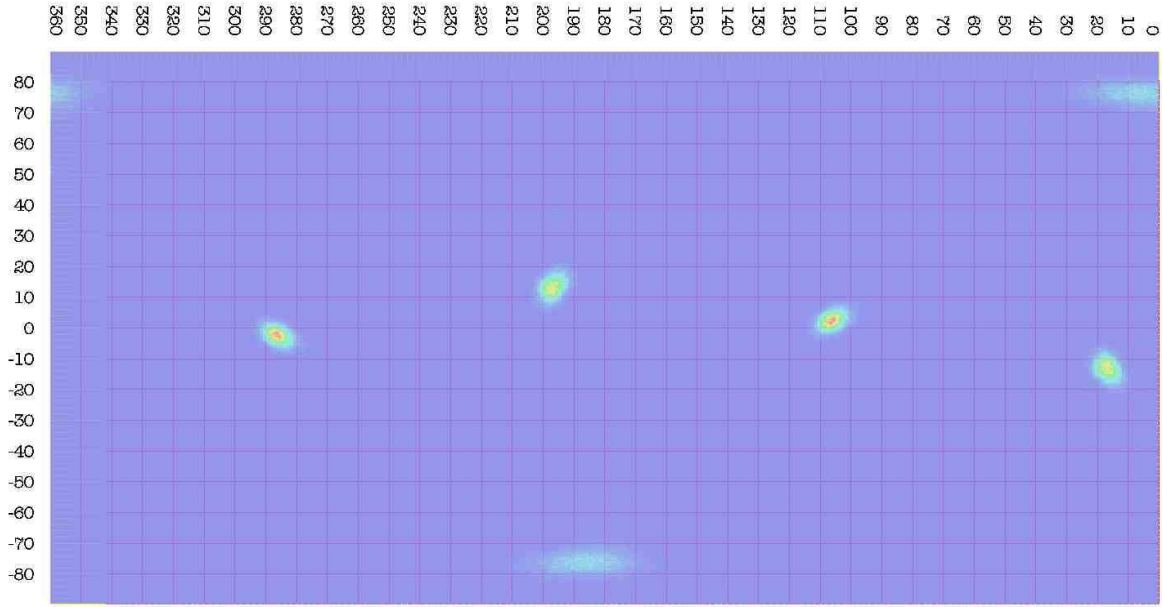}
\end{minipage}
\vspace*{-35pt}
\end{center}
\caption{\label{Fig:mcmc_distribution_angular_gal_coord}
The orientations of the symmetry axes of Fig.\,\ref{Fig:mcmc_distribution_angular}
are shown in Galactic coordinates $(l,b)$.
Note that the meridian $l=0$ goes vertically through the centre of
the Mollweide projection in Fig.\,\ref{Fig:mcmc_distribution_angular}.
}
\end{figure}

\begin{figure}[ttt]
\begin{center}
\hspace*{0pt}\begin{minipage}{12cm}
\vspace*{-5pt}\includegraphics[width=12.0cm]{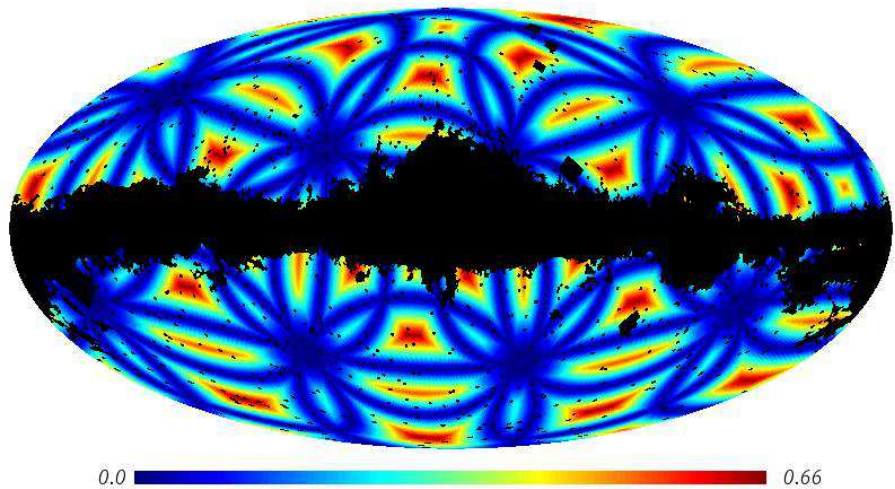}
\end{minipage}
\vspace*{-10pt}
\end{center}
\caption{\label{Fig:MinDist_Torus_ILC_5yr}
The distance $d_{\hbox{\scriptsize min}}(q)$,
eq.\,(\ref{Eq:Distance_min}), is shown for the equilateral toroidal universe
with side length $L=3.84$ and the optimal orientation
shown in Figs.\ \ref{Fig:mcmc_distribution_angular} and
 \ref{Fig:mcmc_distribution_angular_gal_coord}
using the same Mollweide projection as in Fig.\,\ref{Fig:MinDist_Torus}.
The black regions show the pixels excluded by the KQ75 mask.
}
\end{figure}

\begin{table}
\begin{center}
\begin{tabular}{|c|c|c|c|c|c|c|}
\hline $l$ & 6$^\circ$ & 17$^\circ$  & 107$^\circ$  & 186$^\circ$ & 197$^\circ$  &  287$^\circ$\\
\hline $b$ & 77$^\circ$  &  -13$^\circ$  & 3$^\circ$ &  -77$^\circ$  & 13$^\circ$  & 3$^\circ$  \\ 
\hline 
\end{tabular} 
\caption{\label{Tab:Galactic_coordinates}
The mean values of the orientation of the torus in Galactic coordinates $(l,b)$
obtained from the ten Markov chains shown in Fig.\,\ref{Fig:mcmc_distribution_L}.
The accuracy is of order $2^\circ$.
The six directions are either opposite or orthogonal to each other.
}
\end{center}
\end{table}

After having found a preferred region around $L=3.84$,
ten MCMC chains of length 20\,000 are generated which use initial values
$(L_i, \alpha_i, \beta_i, \gamma_i)$ within the likely domain.
These ten chains provide together 200\,000 states from which the
torus orientation and the torus size $L$ is estimated.
Fig.\,\ref{Fig:mcmc_distribution_L} shows the distribution of $L$ for these
Markov chains.
An estimate of $L = 3.84 \pm 0.10$ is obtained.
The directions of the corresponding symmetry axes,
i.\,e.\ the directions to the six centres of the nearest face-to-face neighbours,
are shown in Fig.\,\ref{Fig:mcmc_distribution_angular} in Galactic coordinates
using the same Mollweide projection as in Fig.\,\ref{Fig:MinDist_Torus}.
In Fig.\,\ref{Fig:mcmc_distribution_angular_gal_coord}
these orientations are directly presented in Galactic coordinates
which, however, distort the maxima towards the poles.
Figs.\,\ref{Fig:mcmc_distribution_angular} and
\ref{Fig:mcmc_distribution_angular_gal_coord} reveal the preferred orientation,
whose coordinates are given in Table \ref{Tab:Galactic_coordinates}.
Fig.\,\ref{Fig:MinDist_Torus_ILC_5yr} displays the function
$d_{\hbox{\scriptsize min}}(q)$ for this orientation
as in Fig.\,\ref{Fig:MinDist_Torus} masked by KQ75.
If one uses instead of the KQ75 mask the less restrictive KQ85 mask,
which contains more regions with possible residual foregrounds,
a second slightly shifted orientation arises.
But since the application of the KQ75 mask gives the cleanest pixels 
\cite{Gold_et_al_2008},
the results obtained with this mask are the most reliable.

\begin{figure}[ttt]
\begin{center}
\begin{minipage}{10cm}
\vspace*{-25pt}\includegraphics[width=10.0cm]{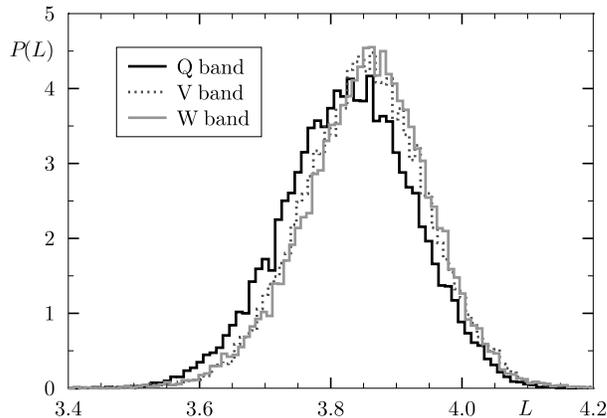}
\end{minipage}
\vspace*{-25pt}
\end{center}
\caption{\label{Fig:mcmc_distribution_L_QVW}
The normalized distribution of the length $L$ is shown according to the states of
five Markov chains obtained from the foreground reduced maps
for each of the bands Q, V, and W using the KQ75 mask.
The likelihood function with $\sigma$, eq.\,(\ref{Eq:sigma_Roukema}),
has been used.
The states of five chains belonging to a given band are merged so
that the histograms are each based on 100\,000 states.
These chains consistently lead to $L_Q = 3.83 \pm 0.10$, $L_V = 3.85 \pm 0.10$,
and $L_W = 3.85 \pm 0.09$ for the Q, V, and W band, respectively.
}
\end{figure}

In order to test whether the spatial-correlation-function signature
is also present in the sky maps of a single frequency band,
five MCMC runs are applied to each of the three bands Q, V, and W
using the KQ75 mask.
For these three bands, foreground reduced maps are available
from the WMAP team \cite{Hinshaw_et_al_2008} based on the
five year data.
It turns out that the MCMC chains,
having again initial values of $L_i$ randomly drawn
from the interval $L_i\in [3.15, 3.45]$,
do indeed find the same orientation and the same length $L$
for the toroidal topology as the chains obtained from the ILC map.
The distributions of the chains are localized in a similar way 
as those obtained from the ILC map
for which an example is shown in Fig.\  \ref{Fig:mcmc_ilc_distribution_L_gamma}.
Using initial points in the high probability domain, further five MCMC chains
with length 20\,000 for each of the three bands are generated.
The distribution of the topological length $L$ derived from these 15 chains
is displayed in Fig.\ \ref{Fig:mcmc_distribution_L_QVW}; and
the obtained values agree well with the result derived from the ILC map.
This shows that four CMB maps, i.\,e.\ the ILC map
and the three foreground reduced maps
for the bands Q, V, and W, give consistent results.

It is worthwhile to remark that the results do not depend on the chosen
probability function, i.\,e.\ using $\sigma$ or $\hat\sigma$ in (\ref{Eq:P_Roukema}).
Five MCMC chains are generated using $\hat\sigma$ for random initial points
in the case of the ILC map using the KQ75 mask, and
the same scale and orientation of the fundamental cell is revealed.
Thus the algorithm is robust with respect to these two probabilities.
Furthermore, the convergence of the Markov chains is checked by
computing the MCMC power spectrum and the convergence ratio $r$
along the lines described in \cite{Dunkley_Bucher_Ferreira_Moodley_Skordis_2005}.

\begin{figure}[ttt]
\begin{center}
\begin{minipage}{10cm}
\vspace*{-25pt}\includegraphics[width=10.0cm]{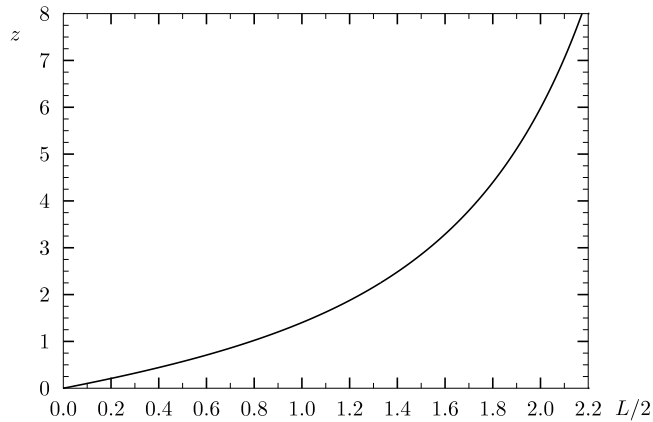}
\end{minipage}
\vspace*{-25pt}
\end{center}
\caption{\label{Fig:length_redshift}
The redshift $z$ at which antipodal objects occur in the direction of the
nearest face-to-face neighbours is shown
in dependence on the distance $L/2$
using the cosmological parameters stated in Section \ref{Sec:MCMC}.
}
\end{figure}

Let us now discuss whether this topology is directly observable
for redshifts below $z=6$.
The topological mirror image of our galaxy cluster is beyond the SLS
because the topological scale $L\simeq 3.8$ is greater than the
distance $\Delta \eta = 3.329$ to the SLS
assuming the cosmological parameters stated in Section \ref{Sec:MCMC}.
The nearest mirror images of objects at the centres of the ``faces'' of the torus,
however, occur at half that length $L/2 \simeq 1.9$ from our point of view
corresponding to a redshift $z\simeq 5.1$,
i.\,e.\ there should be the same objects
at the antipodal points of Table \ref{Tab:Galactic_coordinates} at this redshift.
The strong dependence of this redshift on the distance $L/2$ is shown
in Fig.\,\ref{Fig:length_redshift}.

\section{Discussion}

Does the analysis of the preceding Section indeed point to
a toroidal topology of the Universe
or is the enhanced likelihood around $L\simeq 3.8$ a statistical fluke?
It could be possible that the different contributions to the CMB,
i.\,e.\ mainly the usual Sachs-Wolfe contribution, the Doppler effect,
and the integrated Sachs-Wolfe contribution conspire in such a way
that the cross-correlation function $\xi_{\hbox{\scriptsize C}}(r)$ 
gets by chance relative large values at small distances
for the obtained orientation and torus size.

\begin{figure}[ttt]
\begin{center}
\begin{minipage}{18cm}
\begin{minipage}{9cm}
\includegraphics[width=9.0cm]{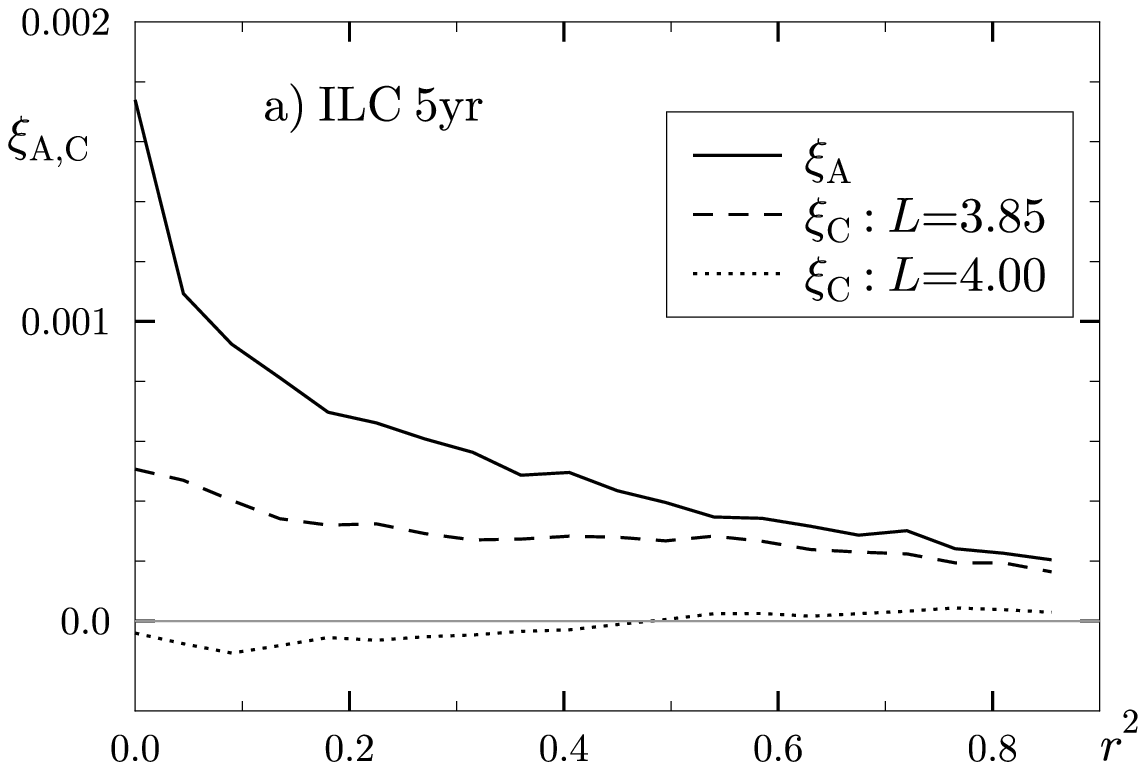}
\end{minipage}
\begin{minipage}{9cm}
\hspace*{-30pt}\includegraphics[width=9.0cm]{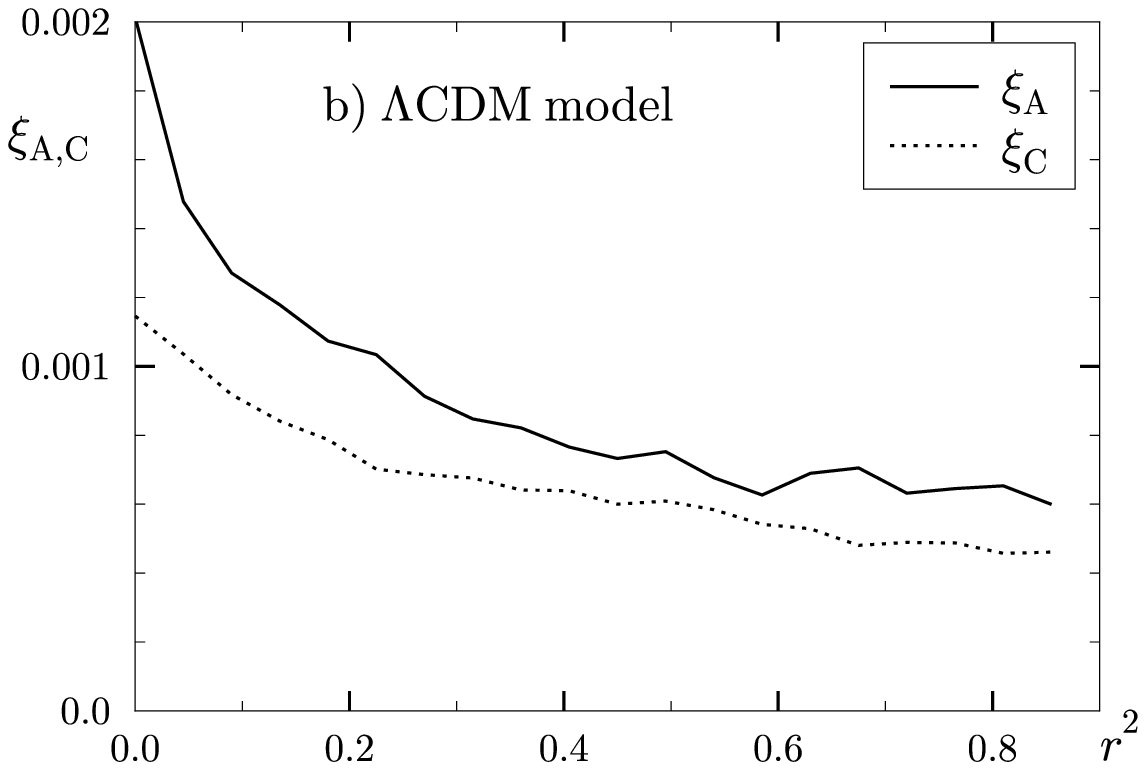}
\end{minipage}
\end{minipage}
\end{center}
\caption{\label{Fig:xi_corr_best}
The cross-correlation $\xi_{\hbox{\scriptsize C}}$ and
the auto-correlation $\xi_{\hbox{\scriptsize A}}$ are shown in panel a)
for two toroidal cells:
one with a correct orientation ($L=3.85$)
and one with a wrong size $L=4.00$ and orientation
$(\alpha=\beta=\gamma=0)$ using the 5yr ILC sky map.
Panel b) displays $\xi_{\hbox{\scriptsize C}}$ and
$\xi_{\hbox{\scriptsize A}}$ for the best toroidal cell for the
"false positive" $\Lambda$CDM model.
The KQ75 mask is applied in all cases.
}
\end{figure}

In order to settle this question one has to generate a large number
of $\Lambda$CDM random sky simulations with the cosmological parameters
of the standard model and apply the algorithm to them.
This would yield the probability by which such a spatial-correlation-function signature
gives a false positive detection.
However, the numerical effort for such an investigation is prohibitive.
Thus, only ten $\Lambda$CDM sky maps are generated here
using the cosmological parameters stated in Section \ref{Sec:MCMC}.
For each of the ten sky maps, five MCMC chains are computed and analyzed.
The KQ75 mask is applied for all sky maps.
No maxima at small values of $L$ are found for nine out of the ten sky simulations
where the probability (\ref{Eq:P_Roukema}) is comparable or even
larger than the one computed from the ILC map.
In these cases the Markov chains are found drifting to ever larger values
of $L$ which means, as discussed above, a decreasing significance for a detection.
However, one of the ten models produces such a nice peak structure 
as seen in Fig.\,\ref{Fig:mcmc_ilc_distribution_L_gamma}
but around $L\simeq 4.17$.
The corresponding cross-correlation $\xi_{\hbox{\scriptsize C}}$ matches
the auto-correlation $\xi_{\hbox{\scriptsize A}}$ even better than the
best match for the torus using the real data
as shown in Fig.\,\ref{Fig:xi_corr_best}.
Furthermore, one of the remaining nine simulations produces a peak structure
with a probability which is a factor 10 lower than that of the ILC map.
Based on these limited observations one might conclude that
$\sim$ 10\% \dots 20\%
of the simulations would produce such a false positive detection.
This is a very high level and thus, the conclusion should rather be
that, if the topology is an equilateral toroidal one, then the torus has probably the
side length $L\simeq 3.85$ and the orientation given in
Tab.\,\ref{Tab:Galactic_coordinates}.

It is intriguing that the two-point temperature correlation function
$C(\vartheta) := \left< \delta T(\hat n)\, \delta T(\hat n') \right>$,
$\hat n\cdot  \hat n' = \cos\vartheta$,
also suggests a torus size around $L\simeq 3.86$ when analyzed with the 3yr ILC map
and applying the kp0 mask \cite{Aurich_Janzer_Lustig_Steiner_2007}.
Evaluating $C(\vartheta)$ without applying any mask leads, however, to
$L\simeq 4.35$.
But nevertheless, both the observed low power at large scales in the CMB 
sky map as well as the spatial-correlation-function signature are compatible
with an equilateral toroidal structure.
However, a firm conclusion cannot be drawn because of the high
false positive rate.

In \cite{Roukema_et_al_2008a} the spatial-correlation-function method is applied
to the Poincar\'e dodecahedral space and also a favoured orientation for
this topology is found.
Of course, only one of both topologies can be realized, if one of them at all.
It is currently not clear which topology gives a better match to the data.
In a following publication this question will be addressed by applying the
spatial-correlation-function signature to several distinct topologies.
Furthermore, it is intended to obtain a better estimate of the false positive rate.


\ack

OpenDX (www.opendx.org), CMBFAST (www.cmbfast.org) and
HEALPix (healpix.jpl. nasa.gov)
\cite{Gorski_Hivon_Banday_Wandelt_Hansen_Reinecke_Bartelmann_2005}
as well as the WMAP data from the LAMBDA website (lambda.gsfc.nasa.gov)
were used in this work.
The computations are carried out on the Baden-W\"urttemberg grid (bwGRiD).

\section*{References}

\bibliography{../bib_astro}

\bibliographystyle{h-physrev3}

\end{document}